\documentclass[12pt,reqno]{article}
\usepackage{amsmath, amsthm, amssymb,epsfig,alltt}

\begin{document}


\title{ Teleportation by a Majorana Medium  }

\author{  Gordon W. Semenoff
\\{\small
Department of Physics and Astronomy, University of
British Columbia,}\\{\small  Vancouver, British Columbia, Canada V6T 1Z1}
\\~~\\Pasquale Sodano\thanks{  Dipartimento di Fisica e Sezione INFN,
 Universita degli Studi di Perugia,
 Via A. Pascoli, 06123, Perugia}\\
{\small Progetto Lagrange, Fondazione C.R.T. e Fondazione I.S.I.}\\
{\small c/o Dipartimento di Fisica- Politecnico di Torino,}\\
{\small Corso Duca degli Abbruzzi 24, 10129,  Torino-Italy} }

\maketitle                 

\abstract{It is argued that Majorana zero modes in a system of
quantum fermions can mediate a teleportation-like process with the
actual transfer of electronic material between well-separated
points.  The problem is formulated in the context of a
quasi-realistic and exactly solvable model of a quantum wire
embedded in a bulk p-wave superconductor. An explicit computation of
the tunneling amplitude is given. }

\vskip 2cm

Teleportation by quantum tunneling\footnote{What is usually referred
  to as teleportation is the transfer of the information that is
  necessary to reconstruct a quantum state\cite{bennet}.  Here, by
  'teleportation' we mean the quantum mechanical tunneling of an
  object over a macroscopic distance.} in one form or another has been
the physicist's dream since the invention of the quantum theory.  The
simplest idea makes use of the fact that the quantum wave-function can
have support in classically forbidden regions and can thus reach
across apparent barriers. Wherever the wave-function has support, the
object whose probability amplitude it describes can in principle be
found.

Of course, the typical profile of a wave-function inside a
forbidden region decays exponentially with distance, so its
amplitude on the other side of that region should be vanishingly
small, particularly if any appreciable distance is involved.  A
slightly more sophisticated idea would be to consider a system
where the wave-function could have support that is peaked in two
spatially separated regions. In that case, one could imagine
populating the quantum state corresponding to that particular
wave-function with an object by interacting with the system in the
region of one of the peaks.  Then, once the state is populated,
the object has non-zero probability of occupying the second peak
and, with some efficiency, which is governed by the statistical
rules of quantum mechanics, could be extracted from the perhaps
far away region of the second peak. It would have been teleported.

This idea could work, however, only if the state described by the
wave-function is well separated from other quantum states that the
object could also take up and which would interfere with this
process. Separation is defined, for example, by considering
stationary states which are eigenstates of a Hamiltonian.  Then it
means that the energy eigenvalue of the state is well separated
from the energy eigenvalues of all of the other states. In fact,
it is this isolation of a two-peaked wave-function that is
difficult to achieve.

To understand the problem, consider a mechanical system of a
quantum particle moving in a potential energy landscape where the
potential has two degenerate minima separated by an energy
barrier. Then, the ground state of the system will have a
wave-function with two peaks, one near each of the minima, and at
least approximately symmetric under exchange of the positions of
the minima.  So, what happens when we attempt to populate the
ground state by interacting with the system in the region of one
of the minima?  Can we place an object into that state and thereby
teleport it to the location of the other wave-function peak at the
other minimum?

The answer seems to be ``No.''. In such a system, there should
always be a second quantum state, nearly degenerate in energy with
the ground state, which is at least approximately antisymmetric in
the positions of the minima. The larger the energy barrier to
regular tunnelling between the minima, the closer to degenerate
are the antisymmetric and symmetric states. When we attempt to
create a particle in the ground state by interacting with the
system at the location of one of the minima, instead of populating
the ground state, we place the particle into a superposition of
the symmetric and anti-symmetric states, the wave-function of
which is localized at the position where we interacted with the
system, and having practically zero amplitude at the second
location. It then proceeds to tunnel to symmetrize its state at
the regular rate for quantum tunnelling, a very slow process in
any system where our semi-classical reasoning is valid and we are
no further ahead.

What we need to find is a quantum system where a quantum state
which is well isolated from other states in the spectrum can have
peaks at different locations.  From the argument above, it will be
difficult to find states of this kind which obey the regular
Schr\"odinger equation. It seems to have a built-in protection
against the sort of non-locality that we are looking for.

Single-particle states that are in some sense isolated are well
known to occur for Dirac equations, particularly when interacting
with various topologically non-trivial background fields such as
solitons, monopoles and instantons. The consequences of fermion
zero modes such as chiral anomalies \cite{Treiman:1986ep} and
fractional fermion number \cite{Jackiw:1975fn},
\cite{Niemi:1984vz} are well known. Consider, for example, the
simple one-dimensional model with Dirac equation $ \left[
i\gamma^\mu \partial_\mu +\phi(x)\right]\psi(x,t)=0 $.  This
describes a fermion moving in one dimension and interacting with a
scalar field $\phi(x)$ which we shall take to have a
soliton-antisoliton profile,
\begin{equation}
\phi(x)= \left\{ \begin{matrix} \phi_0 & x<0 ~,~x>L \cr -\phi_0 &
0<x<L \cr \end{matrix} \right.
\end{equation}
It is easy to see that the equation for energy eigenvalues, which,
choosing a basis for Dirac matrices, we can write as
\begin{equation}\label{dirachamiltonian}
i\left( \begin{matrix} 0 & \frac{d}{dx}+\phi(x)  \cr
  \frac{d}{dx}-\phi(x)& 0 \cr \end{matrix}\right) \left(
\begin{matrix} u_E(x) \cr v_E(x) \cr \end{matrix} \right) = E
\left( \begin{matrix} u_E(x) \cr v_E(x) \cr \end{matrix} \right)
\end{equation} has exactly two bound states with energies and
wave-functions
\begin{eqnarray}
E_+\approx \phi_0
e^{-\phi_0L}~~~ \label{positiveenergystate} \psi_+\approx   \sqrt{\phi_0}
\left(
\begin{matrix}  e^{-\phi_0|x|} \cr - i e^{-\phi_0|L-x|}\cr \end{matrix}
\right) +{\cal O}(e^{-\phi_0L})
\end{eqnarray}
\begin{eqnarray}\label{negativeenergystate}
E_-\approx -\phi_0
e^{-\phi_0L}~~~   \psi_-\approx   \sqrt{\phi_0}
\left(
\begin{matrix}  e^{-\phi_0|x|} \cr  i e^{-\phi_0|L-x|}\cr \end{matrix}
\right) +{\cal O}(e^{-\phi_0L})
\end{eqnarray}
where, sufficient for our purposes, we give only the large $L$
asymptotics -- corrections to all quantities are of higher orders in
$e^{-\phi_0L}$. These states have energy well separated from the
rest of the spectrum, which is continuous and begins at $E=\pm
\phi_0$. $E_\pm$ are exponentially close to zero as $L$ is large.
Each wave-function has two peaks, one near $x=0$ and one near $x=L$.

The ground state of the many fermion system has the negative
energy states filled and the positive energy states empty. A
fermion or anti-fermion are then excited by populating a positive
energy state or de-populating a negative energy state,
respectively. If we create a fermion by populating the positive
energy bound state, it has the wave-function $\psi_+$ given in
eqn.(\ref{positiveenergystate}) which indeed has peaks at both
locations, $x=0$ and $x=L$. Similarly, the anti-fermion has
wave-function $\psi_-$ in eqn.(\ref{negativeenergystate}). Note,
however, that the positive and negative energy states are not
isolated from each other.  They are practically degenerate.   This
degeneracy will prove fatal to our attempt to use these states for
teleportation, for interesting reasons which we shall now relate.
When $\phi_0L$ is large, the energy of the electron and positron
states are almost zero. In this case, one can consider a second
set of almost stationary states which are the superpositions
\begin{eqnarray}\label{symmetricstate}
 \psi_0=\frac{1}{\sqrt{2}}\left(e^{iE_0t}\psi_++e^{-iE_0t}\psi_-\right)
  \approx \sqrt{2\phi_0}\left( \begin{matrix} e^{-\phi_0|x|}\cos E_0t \cr e^{-\phi_0|L-x|}
 \sin E_0t \cr\end{matrix}\right) +{\cal O}(e^{-\phi_0L})
\end{eqnarray}
which, for $t<<\frac{1}{E_0}$, has most of its support near $x=0$ and
\begin{eqnarray}\label{asymmetricstate}
 \psi_L=\frac{1}{\sqrt{2}i}\left(e^{iE_0t}\psi_+ - e^{-iE_0t}\psi_-\right)
  \approx \sqrt{2\phi_0}\left( \begin{matrix} e^{-\phi_0|x|}\sin E_0t \cr -e^{-\phi_0|L-x|}
 \cos E_0t \cr\end{matrix}\right) +{\cal O}(e^{-\phi_0L})
\end{eqnarray}
which has most of its support near $x=L$.  By interacting with the
system at $x=0$, we could as well be dropping the fermion into the
state $\psi_0$, which is localized there and which has
exponentially vanishing probability of occurring at $x=L$ (until
$\sin E_0t$ becomes appreciable, which is just the usual estimate
of tunnelling time through a barrier of height $\phi_0$ and width
$L$).

It might seem bizarre that the relevant state would be anything but
the ground state that has $\psi_-(x)$ populated and $\psi_+(x)$ empty.
It has been shown that as $L\to \infty$, this 'ground state' is an
entangled state of (appropriately defined) fermion
number~\cite{Rajaraman:1982xf,Jackiw:1983uf}. If the system is
prepared in this state, subsequent measurement of the fermion number
of one of the solitons will collapse the wave-function to one where
the fermion, rather than occupying the negative energy state $\psi_-$,
occupies either the state $\psi_{0}$ or the state $\psi_{L}$ which are
localized at $x=0$ and $x=L$, respectively. In these states, the
``fractional fermion number'' of the solitons is a sharp quantum
observable. As seen from the vicinity of each soliton, they are
identical to the Jackiw-Rebbi states~\cite{Jackiw:1975fn} of the
fermion in a single soliton background field, which have fermion
number $\pm \frac{1}{2}$. This issue has recently been
re-examined~\cite{Jackiw:2000jx} in conjunction with some ideas about
entangled electron states in Helium bubbles~\cite{Maris}.

We would now imagine that our dumping the fermion into the bound
state, if performed near $x=0$ would populate the state $\psi_0$,
rather than $\psi_+$, as this is the state with sharp local
fermion number and it would have appreciable probability of
appearing at $x=L$ only after a time over order $\frac{1}{\phi_0}e^{\phi_0L}$. Our
quest for a teleportation device has been foiled again by the
existence of degenerate states, this time the slightly more subtle
case of a fermion and anti-fermion state.

The situation is somewhat better if we consider a different type of
fermion, called a Majorana fermion.  The Hamiltonian of a Majorana
fermion has a symmetry which maps positive energy states onto negative
energy states. In the case of (\ref{dirachamiltonian}), we have $
\psi_{-E}(x)= \psi^*_{E}(x)$.  Then, a fermion and an anti-fermion
have the same spectrum, and we simply identify them as the same
particle, a Majorana fermion. The fermion no longer has a conserved
total fermion number. However, it still has fermion parity: fermion
number conservation mod 2.  In any process, Majorana fermions must be
created or annihilated in pairs. This means that we should be able to
classify all quantum states by eigenvalues of an operator $(-1)^F$,
such that a state with an even number of fermions has $(-1)^F=1$
whereas a state with an odd number has $(-1)^F=-1$. This
classification of states has been argued to be of fundamental
importance in three-dimensional quantum
physics~\cite{Streater:1989vi,Andreev:2003nf}. Even though our present
example is one-dimensional, we can imagine that it is an effective
theory which is embedded in the three dimensional world.  The example
we will discuss later is of this sort.

Now, for the Majorana fermion, there is only one bound state, the wave-function
of which is $\psi_+$ (the complex conjugate of which is $\psi_-$),
and that bound state can be either occupied or empty. We can   assign
$(-1)^F=1$ for the empty state and $(-1)^F=-1$ for the
occupied state.
The states corresponding to $\psi_0$ and $\psi_L$ do not have
a definite fermion parity. If we begin with the
system where the quantum state is an eigenstate of fermion parity
and we by some process dump a fermion into the bound state near
$x=0$, its wave-function automatically has a second peak at $x=L$
and it could in principle be extracted there. This is what we mean
by ``teleportation''.  We will argue later  that, locally, if we
live near $x=0$ and are unaware of the region near $x=L$,
a vestige of this phenomenon will appear as either violation of conservation of
fermion parity or the existence of a hidden variable in the
local theory.
Similar peculiar phenomena involving Majorana zero modes have previously
been discussed in the context of
supersymmetric field theories with
solitons~\cite{Losev:2000mm,Losev:2001uc}.

Unfortunately, Majorana fermions are not easy to come by in
nature.  The electron is a complex fermion.  Ordinarily, if we
decompose it into its real and imaginary parts, which would be
Majorana fermions, they are rapidly re-mixed by electromagnetic
interactions. One place where this decomposition is done for us
more efficiently is in a superconductor where, because electric
charges are efficiently screened, the Bogoliubov quasi-fermions
behave as if they are neutral excitations. These quasi-particles
can be Majorana fermions when the super-conducting condensate is
parity odd, P-wave being the simplest example~\cite{an}. In these materials,
another common occurrence are mid-gap bound states, the analog of
our fermion zero modes, called Andreev states \cite{and}, which typically
live at surface of the superconductor.
  Majorana zero modes of the type that we are
discussing are also known to be bound to vortices in p-wave
superconductors where they have the remarkable effect of giving
vortices non-Abelian fractional
statistics~\cite{Read}-\cite{stone}. For concreteness we will
consider a slightly simpler system that was originally discussed by
Kitaev~\cite{kitaev}. We consider a quantum wire embedded in a
bulk P-wave superconductor, depicted in Fig.1.
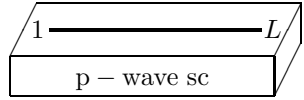
\begin{figure}[ht] \hbox to \textwidth{\hfill
\begin{picture}(110,35)
\put(0,0){\line(1,0){100}} \put(0,15){\line(1,0){100}}
\put(10,35){\line(1,0){100}} \put(0,0){\line(0,1){15}}
\put(100,0){\line(0,1){15}} \put(110,20){\line(0,1){15}}
\put(0,15){\line(1,2){10}} \put(100,15){\line(1,2){10}}
\put(100,0){\line(1,2){10}}
\put(15,24.7){\thicklines\line(1,0){80}}
\put(15,25.3){\thicklines\line(1,0){80}}
\put(8,22){\footnotesize$1$} \put(96,22){\footnotesize$L$}
\put(0,4){\hbox to 100\unitlength {\footnotesize\hfil${\rm p-wave
~sc}$\hfil}}
\end{picture}
\hfill} \caption{A  quantum wire on a bulk P-wave superconductor.}
\label{fig_layout}
\end{figure}
We assume that the wire has a single channel and that the
electrons are adequately described by a tight-binding model.  We
also assume a weak coupling to the superconductor whereby the
electrons can enter and leave the wire as Cooper pairs. The
Hamiltonian is
\begin{equation} \label{ham}
H=\sum_{n=1}^L \left( \frac{t}{2}a_{n+1}^\dagger
a_n+\frac{t^*}{2}a_n^\dagger a_{n+1}+\frac{\Delta}{2}
a^{\dagger}_{n+1}a^{\dagger}_n + \frac{\Delta^*}{2} a_n
a_{n+1}+\mu a_n^\dagger a_n\right)
\end{equation}
The first terms in the Hamiltonian are the contribution of hopping
of electrons between sites, labelled $n=1,2,...,L$, on the quantum
wire, where $t$ is the hopping amplitude and
$\left\{a_m,a_n^\dagger\right\}=\delta_{mn}$ are annihilation and
creation operators for electrons. The second pair of terms arise
from the presence of the super-conducting environment. The last
term is the energy of an electron sitting on a site of the wire.
We shall assume that $|\mu|<|t|$ and $|\Delta|<|t|$.

For simplicity, we shall consider spinless fermions. This is possible when the electron spectrum
is strongly polarized so that the energy spectra of spin up and spin down electrons are well
separated.  This is a common occurrence in anisotropic
superconductors~\cite{an}.  In particular, the Cooper pair can be a bound state of spin up electrons only:
in that case the terms with $\Delta$ in (\ref{ham}) which take into account that electron on the quantum
wire can form Cooper pairs would only couple to the spin up electron. The other spin state would behave as a spectator
and to the extent that it would be coupled to the spin up state, it
could be taken into account by an effective Hamiltonian.  We have not confirmed that this would not destroy the unpaired
Majorana zero mode that we shall find, but it is quite plausible that, at least for some range of parameters, the zero mode and teleportation
will persist.

If $t=|t|e^{i\phi}$ and $\Delta = |\Delta|e^{2i\theta}$, by
redefining $a_n\to e^{i(\phi+\theta)}a_n$ for $n$ odd and $a_n\to
e^{i(\phi-\theta)}a_n$ for $n$ even, we remove the complex phases
of $t$ and $\Delta$, which we henceforth assume to be positive
real numbers. The equation of motion for the fermion is
$$
i\frac{d}{dt} a_n =
\frac{t}{2}\left(a_{n+1}+a_{n-1}\right)-\frac{\Delta}{2}\left(a_{n+1}^\dagger
- a_{n-1}^\dagger\right) + \mu a_n
$$
We decompose the fermion into real and imaginary parts,
$a_n=b_n+ic_n$, and assemble them into a spinor $\psi_n = \left(
\begin{matrix} b_n \cr c_n \cr \end{matrix} \right) $ for which we make the ansatz
$\psi_n(t) = e^{i\omega t}\psi_n$.   It obeys the difference
equation \begin{equation}\label{diff} \psi_{n+2}=-N\psi_{n+1} - M
\psi_n \end{equation} where
$$
M=\left[ \begin{matrix} \frac{t+\Delta}{t-\Delta} & 0 \cr 0 &
\frac{t-\Delta}{t+\Delta} \cr \end{matrix} \right] ~~,~~ N =
\left[ \begin{matrix} \frac{2\mu}{t-\Delta} &
\frac{2i\omega}{t-\Delta} \cr \frac{-2i\omega}{t+\Delta} &
\frac{2\mu}{t+\Delta} \cr \end{matrix} \right]
$$

To solve (\ref{diff}) we define a generating function from which
we can recover the wave-function by doing a contour integral
\begin{equation}\label{generating} \psi(\zeta) = \sum_{n=1}^L \zeta^n\psi_n ~~,~~
 \psi_n =
\oint_C \frac{d\zeta}{2\pi i}\frac{1}{\zeta^{n+1}}\psi(\zeta)
\end{equation} where $C$ is a contour of infinitesimal radius
encircling the origin. We will consider a semi-infinite wire by
putting $L\to \infty$.  In this case, the problem is exactly
solvable. The generating function is easily obtained from
(\ref{diff}) as
\begin{equation}\label{solution}
\psi(\zeta) = \frac{\zeta} { 1 +N\zeta +M\zeta^2}\psi_1 =\frac{
\Omega(\theta) }{ \left[ t\cosh\theta +\mu\right]^2-\left[
\Delta\sinh\theta\right]^2-\omega^2 } \tilde\psi_1
\end{equation}
where $\zeta = e^{\theta}$, $\Omega(\theta)=\left[\begin{matrix}
t\cosh\theta+\mu -\Delta\sinh\theta & -i\omega \cr i\omega &
t\cosh\theta+\mu +\Delta\sinh\theta \cr \end{matrix}\right]$ and
$\tilde\psi_1= \left[\begin{matrix} \frac{t-\Delta}{2} & 0\cr 0 &
\frac{t+\Delta}{2}\cr
\end{matrix}\right]\psi_1$  .
We will do the integral in (\ref{generating}) by inverting the
contour, using the fact that the integrand vanishes at infinity so
that the contour can be closed on the poles of $\psi(\zeta)$.  For
generic real values of the frequency $\omega$, $\psi(\zeta)$   has
four poles which solve a quartic equation $\left[ t\cosh\theta_i
+\mu\right]^2-\left[ \Delta\sinh\theta_i\right]^2=\omega^2 $,
which is also the dispersion relation. The integral is then given
by the sum of four residues
\begin{equation}\label{wvfn}
\psi_n =- \sum_{i=1}^4 e^{-n\theta_i}\left\{
\frac{4e^{\theta_i}}{t^2-\Delta^2}\frac{1 }{\prod_{j\neq
i}\left(e^{\theta_i} -
e^{\theta_j}\right)}\right\}\Omega(\theta_i) \tilde\psi_1
\end{equation}
For allowed values of $\omega$, these are the (un-normalized)
wave-function of the quasi-particles.  The physically allowed
values of $\omega$ are determined by requiring normalizability of
the wave-function.

There is a region of continuum spectrum where $\theta_i$ are
imaginary and
$$\omega(k)=\pm\sqrt{\left[ t\cos k +\mu\right]^2+\left[
\Delta\sin k\right]^2}~~,~~k\in(-\pi,\pi]$$ For $\omega$ in this
range, the four roots are $\theta_i=(ik_+,-ik_+,ik_-,-ik_-)$ which
satisfy $$ k_{\pm} = \arccos\left[-\frac{\mu
t}{t^2-\Delta^2}\pm\sqrt{\left( \frac{\mu
t}{t^2-\Delta^2}\right)^2-\left(
\frac{\mu^2+\Delta^2-\omega^2}{t^2-\Delta^2}\right)} \right]$$
They must be used in (\ref{wvfn}) to get the continuum
wave-function with energy $\omega$.

When $\omega$ is not in the range of continuum spectrum, the roots
obey

$$\cosh\theta = \frac{-\mu t\pm i
  \sqrt{   \Delta^2 \left(t^2-\Delta^2-\mu^2\right) -
\omega^2\left(t^2-\Delta^2\right)}}{t^2-\Delta^2}
$$
and two of them, say $e^\theta$ and $e^{\theta^*}$, must have
moduli less than one, leading to a potential diverging part in the
wave-function. The divergence can only be avoided if the spinor
$\tilde \psi_1$ is chosen as a simultaneous zero eigenvector of
the two matrices $\Omega(\theta)$ and $\Omega(\theta^*)$. Recall
that the dispersion relation ensures that these matrices have
vanishing determinant, so each has a vanishing eigenvalue.  They
can have a simultaneous vanishing eigenvalue only if they commute,
$$
\left[ \Omega(\theta),\Omega(\theta^*)\right]=i\omega\left(
\sinh\theta - \sinh\theta^*\right)\left(\begin{matrix} 0&1\cr
1&0\cr\end{matrix}\right)
$$
Since in the range of frequencies of interest, $\sinh\theta$
cannot be real, the only possible bound states are zero modes,
$\omega=0$. By appropriate choice of the spinor, $\tilde\psi_1$ we
find the zero mode wave-function is ~~$b_n^0=0$
\begin{equation}
c_n^0 = \frac{i \left[ \left(-\mu + i\sqrt{t^2-\Delta^2-\mu^2
  } \right)^n
-\left(  -\mu  - i\sqrt{ t^2-\Delta^2-\mu^2  } \right)^n\right]}
{\sqrt{ 2\frac {t(t^2-\Delta^2-\mu^2) }{ \Delta(t^2-\mu^2)}}(t+\Delta)^n}
\end{equation}
There is exactly one bound state of one of the real components of
the electron living at the edge of the quantum wire.  Its energy
is separated from the rest of the quasiparticles which are in the
continuum spectrum.

Note that the existence of a mid-gap state for a Majorana fermion is
quite robust.  Since, in order to have a Majorana fermion in the first
place, there must exist a one-to-one mapping between positive and
negative energy states of the single-fermion Hamiltonian, smooth
perturbations of it can only change the number of zero modes by an
even integer.  Generically, such perturbations would lift the zero
modes, leaving the one that we have found.

Let us assume that the wire is effectively semi-infinite in the
sense that we are only aware of the edge at $n=1$.  We would second
quantize the fermions as
\begin{eqnarray}\label{second1}
b_n(t)= \int_{-\pi}^\pi\frac{dk}{2\pi}\left(
b_n(k)e^{i\omega_kt}\alpha_k+ b^*_n(k)e^{-i\omega_kt}
\alpha^\dagger_k \right)
\\ \label{second2}
c_n(t)= c^0_n \beta^0+\int_{-\pi}^\pi\frac{dk}{2\pi}\left(
c_n(k)e^{i\omega_kt}\beta_k+
c^*_n(k)e^{-i\omega_kt}\beta^\dagger_k \right)
\end{eqnarray}
where we now assume that all wave-functions are normalized and the
creation and annihilation operators for quasiparticles obey the
usual anti-commutator algebra  $$\left\{\alpha_k,
\alpha^\dagger_{k'}\right\} =\delta(k-k')~~,~~\left\{\beta_k,
\beta^\dagger_{k'}\right\} =\delta(k-k')~~,~~\left\{\beta_k,
\alpha_{k'}\right\} =0~.
$$
The zero mode operator obeys $(\beta^0)^2=1$ and anti-commutes
with all other operators. A representation of the
anticommutator algebra begins with an eigenstate of $\beta^0$,
\begin{equation}\label{plus}
\beta^0 |+> =|+>
\end{equation}
which is annihilated by all other operators $\alpha_k |+> =0=
\beta_k|+>$.  Quasi-particle states are then given by creation
operators acting on this ground state,
$\alpha_{k_1}^\dagger...\beta_{{k_1}'}^\dagger...|+>$.

This is an irreducible representation of the creation and
annihilation operator algebra and is what we would use if we were
unaware of the other end of the wire. The ground state $|+>$ is
not an eigenstate of fermion parity, so fermion
number mod 2 is not conserved in this system. We could then imagine a
process where we probe the system with a source which couples to
electric charge density, for example $$H_{\rm int}=\sum_n
V(n,t)a_n^{\dagger}(t)a_n(t).$$ Recalling that
$a_n(t)=\frac{1}{\sqrt{2}}\left( b_n+ic_n\right)$ and using
(\ref{second1}) and (\ref{second2}), see that there is a finite
amplitude for the annihilation of a single fermionic quasi-particle.
\begin{equation}\label{vanish}
<+|~ H_{\rm int}~ \alpha_k^{\dagger}|+> = ie^{i\omega_k t}\cdot
\sum_n \left(V(n,t)b_n(k)c_n^0\right)
\end{equation}
It literally looks like the fermion vanished while traversing the
region near the edge of the wire.  The inverse process of
making a quasi-fermion appear has an amplitude which is the
complex conjugate of the above.

Of course, the total system that we are considering does have
fermion parity symmetry.  The Hamiltonian is quadratic in fermion
operators, as is the probe that we are using.  The symmetry is being broken by our
insistence on the choice of an irreducible representation of the
anti-commutator algebra.  It can be restored by using a reducible
representation instead.  The minimal such restoration is equivalent
to introducing a single hidden variable.

If the system has another boundary at large but finite $L$, the most
significant effect is that the bound state is not longer an exact zero
mode. Now there are a pair of states with positive and negative
energies of order $\pm\Delta e^{-\Delta L}$.  We shall assume that $L$
is large enough that these energies are negligibly small.  Similar to
the relativistic case that we studied earlier
(\ref{positiveenergystate}) and (\ref{negativeenergystate}), the
wave-functions of each of these bound states has support at both ends
of the wire and is vanishingly small in between.  When $L$ is large,
the rest of the spectrum resembles the continuum spectrum of the
semi-infinite wire.  Now, second quantization introduces two operators
corresponding to the (almost) zero modes, rather than the one which
was found for the semi-infinite wire: we can think of it as producing
the hidden variable which we discussed above.

The zero mode localized near $x=L$ will be a mode of $b_n(t)$ with
wave-function $b_n^0 = c^0_{L+1-n}$. To second quantize the Majorana
fermions, we modify (\ref{second1}) as
\begin{eqnarray}\label{second3}
b_n(t)= c_{\tiny L+1-n}^0\alpha^0+\int_{-\pi}^\pi\frac{dk}{2\pi}\left(
b_n(k)e^{i\omega_kt}\alpha_k+ b^*_n(k)e^{-i\omega_kt}
\alpha^\dagger_k \right)
\\ \label{second4}
c_n(t)= c^0_n \beta^0+\int_{-\pi}^\pi\frac{dk}{2\pi}\left(
c_n(k)e^{i\omega_kt}\beta_k+
c^*_n(k)e^{-i\omega_kt}\beta^\dagger_k \right)
\end{eqnarray}
To represent the anti-commutator algebra we now need the second
eigenstate of $\beta^0$, $\beta^0|->=-|->$. Since $\alpha^0$ must
obey $\left\{\alpha^0,\beta^0\right\}=0$ and
$\left(\alpha^0\right)^2=1$, its action on these states must be
(up to choice of phase)
$$
\alpha^0 |+> = |-> ~~,~~ \alpha^0 |-> = |+>
$$
Now, we have two degenerate ground states $|+>$ and $|->$.  We can
in fact find  states which are eigenstates of fermion parity by
taking superpositions
$$
|\uparrow>=\frac{1}{\sqrt{2}}\left(|+>+i|->\right) ~~,~~
|\downarrow>=\frac{1}{\sqrt{2}}\left(|+>-i|->\right)
$$
In this space $\beta^0|\uparrow>=|\downarrow>$, and
$\beta^0|\downarrow>=|\uparrow>$, $\alpha^0|\uparrow>
=i|\downarrow>$ and $\alpha^0|\downarrow>=-i|\uparrow>$. We can choose $|\uparrow>$ and $|\downarrow>$ to be
eigenstates of fermion parity, $ (-1)^F|\uparrow>=|\uparrow>$,
$(-1)^F|\downarrow>=-|\downarrow> $.

The process analagous to the one in (\ref{vanish}) has the
non-zero matrix element
\begin{equation}\label{vanish2}
<\uparrow|  ~ H_{\rm int}~\alpha_k^\dagger  |\downarrow> =
ie^{i\omega_k t}\cdot \sum_n \left(V(n,t)b_n(k)c_n^0\right)
\end{equation}
and the destruction of a single quasi-fermion is accompanied by a
flip in the vacuum state, the total process conserving $(-1)^F$.

What about teleportation?  Let us imagine that we begin with the
system in one of its ground states and inject an electron at site
$\# 1$, to create  the state $a_1^{\dagger}|\uparrow>$. We then
ask what is the quantum transition amplitude for the transition,
after a time $T$, of this state to one with the electron located
at position $\# L$, $a_L^{\dagger}|\uparrow>$.  The amplitude is
given by
\begin{equation} {\cal A}_{1L} =\frac{ <\uparrow|~a_L ~e^{iHT}~
a_1^{\dagger}~|\uparrow>}{|a_L^\dagger|\uparrow>|~
|a_1^\dagger|\uparrow>|} = 2|c_1^0|^2+\left(T{\rm ~ and~}
L{\rm-dependent} \right)
\end{equation}
The right-hand-side of this equation has two kinds of terms.
The $T$- and $L$-dependent parts of this matrix element represent
the usual propagation via excited quasi-particles which must
travel across the wire. The first term, on the other hand, is unusual
in that it is $T$ and
$L$-independent. By `teleportation', we are referring to this part
of the amplitude.
The probability that the electron, after being injected at site $\# 1$
is teleported to any site $\# n$ within the
support of the zero mode at $n=L$ is
\begin{equation} P=\sum_n|{\cal A}_{1n}|^2 = 4
\frac{\Delta}{t}\frac{t^2-\mu^2}{(t+\Delta)^2}
\end{equation}
which can be significant.

Note that this tunneling process does not violate causality.  There is
a finite probablility for an electron to appear at site $\# L$
spontaneously accompanied by a flip between the ground states, $$
<\uparrow|a_Le^{iHT}|\downarrow> \neq 0 $$
In order for an observer at
site $\# L$ to know that the electron arriving there was not produced
by this spontaneous flip, the information about the state in which the
system was initially prepared would have to be sent to her/him by
independent means.

We note that, if the fermion parity superselection rule is respected,
the degenerate ground states behave like a classical switch.  The
switch can be in either of two positions $|\uparrow>$ or
$|\downarrow>$ but a quantum superposition of the two states is not
allowed. The quantum coherence of this system leads to this surprising
classical behavior.

This has a further interesting effect when one considers two such
systems.  This consideration is in fact natural if there is spin
degeneracy.  For simplicity, we consider two spin states that are
sufficiently weakly coupled to each other that they satisfy
independent single-fermion Schroedinger equations and the quantum
states are direct products of the states of the two systems.  Then,
the quantum states that are allowed by the fermion parity
superselection rule cover two Bloch spheres, for the separate cases
where fermion parity is odd or even, respectively $$|\theta\phi>=
\cos\frac{\theta}{2} |\uparrow\downarrow>+
e^{i\phi}\sin\frac{\theta}{2} |\downarrow \uparrow>
,~~~|\theta'\phi'>= \cos\frac{\theta'}{2}
|\uparrow\uparrow>+e^{i\phi'}\sin\frac{\theta'}{2}|\downarrow\downarrow>$$
The expectation value of the spin vanishes in both of these states,
independently of the angles.  This means that the ground state
degeneracy is not split by a sufficiently weak Zeeman interaction, in
line with the fact that the existence of the Majorana zero modes is
independent of the interactions -- the Hamiltonians for the two spin
species need not be identical, they only need to each have the
conditions for existence of the Majorana zero mode in the first place.
In the present case, the energy splitting of the spin states due to a
Zeeman interaction with an external magnetic field would simply be
carried by a relative shift of chemical potential $\mu$ for each spin
which, if small enough, leaves the zero mode sector intact.  The
amplitude that we have computed for tunneling into zero modes in this
case is just a sum of the amplitudes for each spin state.  The states
of a given fermion parity form a qubit whose properties are very
stable and which could be of use for quantum computation.

\vskip .5cm

\noindent G.S. thanks A.Kitaev, A.Leggett, G.Milburn and W.Unruh for
discussions and acknowledges the hospitality of the I.H.E.S.,
Bures-sur-Yvette, the Werner Heisenberg Institute, Munich and the
University of Perugia.  P.S. thanks M.Rasetti for discussions and
acknowledges the Pacific Institute for Theoretical Physics, Vancouver
and the Center for Theoretical Physics at MIT for hospitality. This
work is supported in part by NSERC of Canada and by M.I.U.R.~National
Project ''JOSNET'' (grant n.2004027555).

\end{document}